\documentclass[onecolumn,letterpaper,12pt]{article}

\usepackage{amssymb}
\usepackage{amsmath}
\usepackage{graphicx}

\begin{document}


\title{Visualizing long vectors of measurements by use of the Hilbert curve}

\author{E. Estevez-Rams\\
Facultad de F\'isica-IMRE, \\
University of Havana, San Lazaro y L. CP 10400.
La Habana. Cuba.\\
Corresponding author. e-mail: estevez@imre.oc.uh.cu\\
\and
C. Perez-Davidenko\\
Instituto de Ciencias y Tecnolog\'ia de Materiales(IMRE),\\
University of Havana, San Lazaro y L. CP 10400.
La Habana. Cuba.\\
\and
B. Arag\'on Fern\'andez\\
Universidad de las Ciencias Inform\'aticas (UCI),\\
Carretera a San Antonio, Boyeros. La Habana. Cuba.\\
\and
R. Lora-Serrano\\
Universidade Federal de Uberlandia, AV. \\
Joao Naves de Avila, 2121- Campus Santa Monica, CEP 38408-144,\\
Minas Gerais, Brasil.
}


\date{\today}
\maketitle

\begin{abstract}
The use of Hilbert curves to visualiza massive vector of data is revisited following previous authors. The Hilbert curve mapping preserves locality and makes meaningful representation of the data. We call such visualization as Hilbert plots.  The combination of a Hilbert plot with its Fourier transform allows to identify patterns in the underlying data sequence. The use of different granularity representation also allows to identify periodic intervals within the data. Data from different sources are presented: periodic, aperiodic, logistic map and 1/2-Ising model. A real data example from the study of heartbeat data is also discussed. 
\end{abstract}

\section{Introduction}

A typical trait in many fields of science is the existence of a large amount of data resulting either from experimental measurements or from some other sources such as computer simulations. An archetypical example is that of a long time series, where the whole data set can be thought as a very large vector. Many combinatorial, statistical, informational theory tools among others have been developed  to recognize or discover peculiar features in such data vectors. Common tasks can involve exploring regularities, extracting periodic intervals, asserting  ''randomness'', among others. In spite of the availability of such powerful quantitative tools, the need of displaying the whole set of values in a meaningful way becomes relevant. The visual inspection of the data can help in identifying peculiarities and in discovering structures from where hypothesis can be advanced or decisions can be made, as to which tools are the most convenient to be applied to the actual data series. In a general setting, the one dimensional sequence is to long to make sense writing down all values of the string. The local inspection of the string of values can be useful in some contexts, but it has the serious shortcoming that the overall picture of the data can be missed. In local exploration, although features occurring in short periods can be identified, the logical arrangement of such features over the whole dataset can remain hidden to the observer. Previous authors have used the good locality preserving qualities of Hilbert curve for the purposes of data visualization in general \cite{keim96} and in the context of DNA analysis \cite{anders09}. Following such ideas, in this contribution we show the use of the Hilbert curve for displaying data vectors in a meaningful way for synthetic and real life data relevant to physics. We call such representation Hilbert plots and they display the whole of the data vector in a two dimensional (2D) plot that does not hinder the inspection of local intervals if needed. Furthermore, it will be shown that Hilbert plots combined with its Fourier transform, can be powerful enough to carry out some of the uses we have listed above to be expected from data visualization. 

If we are seeking for some kind of optimality in visualizing datasets, neighborhood preserving and clustering are two good criteria for a large number of applications. Among the different one-to-one mapping between n-dimensional and one-dimensional (1D) spaces, Hilbert curve has the best neighborhood preserving properties \cite{bauman06,moon01,jagadish97}. It is a special case of space filling mapping known as Peano curves,  that allows to map a 1D sequence into a multidimensional space and viceversa. The use of Hilbert curve  has been extensively studied and used in a wide range of applications \cite{cox94,songa02,liang08,anders09,chen11}.

The display of one dimensional data series as 2D ``images'' is not new. A 'logical' approach would be the sequential partition of the data sequence in equal segments of values in order to build a display matrix using shades of gray or any other coloring scheme. Such approach has the serious drawback that far apart values in the original series, will end up close in the two dimensional plot, giving rise to false neighbors. The artificial creation of false neighborhoods could lead to unfounded conclusions from the visual inspection of the plot. As serious as the former problem, the visualization devised in such a way, do not have nice clustering properties, and a substring of equal values will be displayed as a single line with the same color, hard to distinguish in a large array.

The remainder of the article is organized as follows. In section \ref{sec:pre} some notation and preliminary notions will be introduced. Granularity will be defined, as well as the dilation factor as a measure of locality preservation.  The row scan will be described. In section \ref{sec:hc} we define the Hilbert curve construction, and analyze some of its properties. Hilbert plots will then be defined. In section \ref{sec:cases} we discuss the use of Hilbert plots for visualizing pure periodic data and interleaved periodic sequences; aperiodic data as those resulting from fixed morphism (such as Thue-Morse and Fibonacci sequences); data from the binary partition of non-linear maps such as the logistic map; and finally, data from the 1/2-Ising model with first neighborhood interactions. Conclusions follow.

\section{Preliminaries}\label{sec:pre}

\begin{figure}
\includegraphics[scale=1.3]{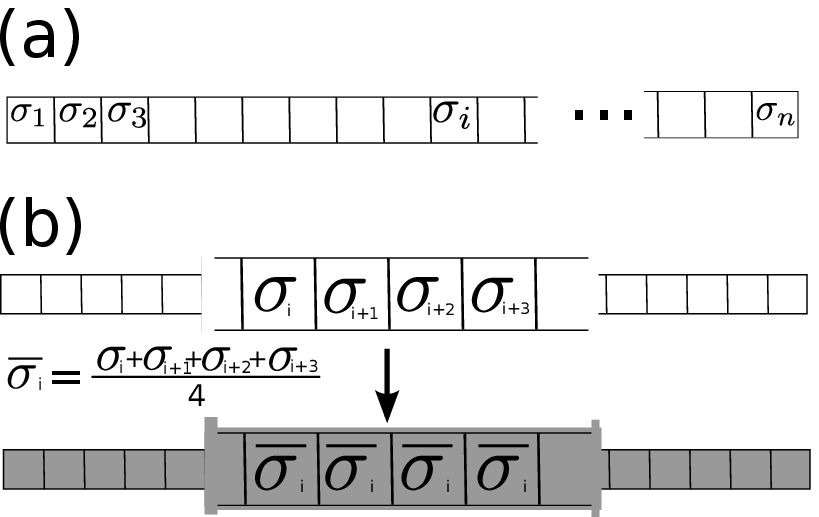}
\caption{A data sequence as a vector (a) of values $\sigma_{i}$ indexed from left to right. (b) The mean value
approximation of 4-granularity is constructed by a sliding window of length 4, where each non-overlapping partition
$\sigma_{i;4}$ is substituted by a constant subsequence of the same length and value $\overline \sigma_{i}.$
}\label{fig:sigma}
\end{figure}

Consider a sequence of values organized in a one dimensional string $\varSigma$ of length $n=|\varSigma|$. Each value $\sigma_{i}$ will be indexed by $i$, giving its position in the string from left to right (Figure \ref{fig:sigma}a). The distance between two positions $i$ and $j$ in the string will be the absolute value of the difference between their indexes $|i-j|$. A substring of length $l$ and starting at index $i$  will be denoted by $\sigma_{i;l}$ ($\equiv\sigma_{i}\sigma_{i+1}\ldots\sigma_{i+l-1}$).

In what follows it will be important to define granularity over a sequence $\varSigma$. The $l$-granularity representation of string $\varSigma$, is the mean approximation over substrings of length $l$. If  a partition of the string $\varSigma$ into non overlapping substrings $\sigma_{i;l}$ is performed, then for each substring a mean value can be calculated:

\begin{equation}
 \overline{\sigma_{i}^{l}}=\frac{1}{l}\sum_{j=0}^{l-1}\sigma_{i+j},
\end{equation}

and the following substitution is made (Figure \ref{fig:sigma}b) 

\begin{equation}
 \sigma_{i;l} \longrightarrow \overline{\sigma}_{i;l},
\end{equation}

where

\begin{equation}
 \overline{\sigma}_{i;l}=\overline{\sigma_{i}^{l}}\overline{\sigma_{i}^{l}}\overline{\sigma_{i}^{l}}\ldots\overline{
\sigma_ { i } ^{l}}
\end{equation}

is the mean approximation of the original substring. 

$1$-granularity representation of the sequence is, by construction, equal to the original $\varSigma$, which will also be called the faithful representation. For $l>1$, the $l$-granularity of a sequence gives a  coarser view of $\varSigma$.  For a periodic sequence of period $p$, the $lp$-granularity representation of $\varSigma$ ($l$ a natural number) will result in a constant vector with components having the mean value over the periodic unit.

\begin{figure}
\includegraphics[scale=0.75]{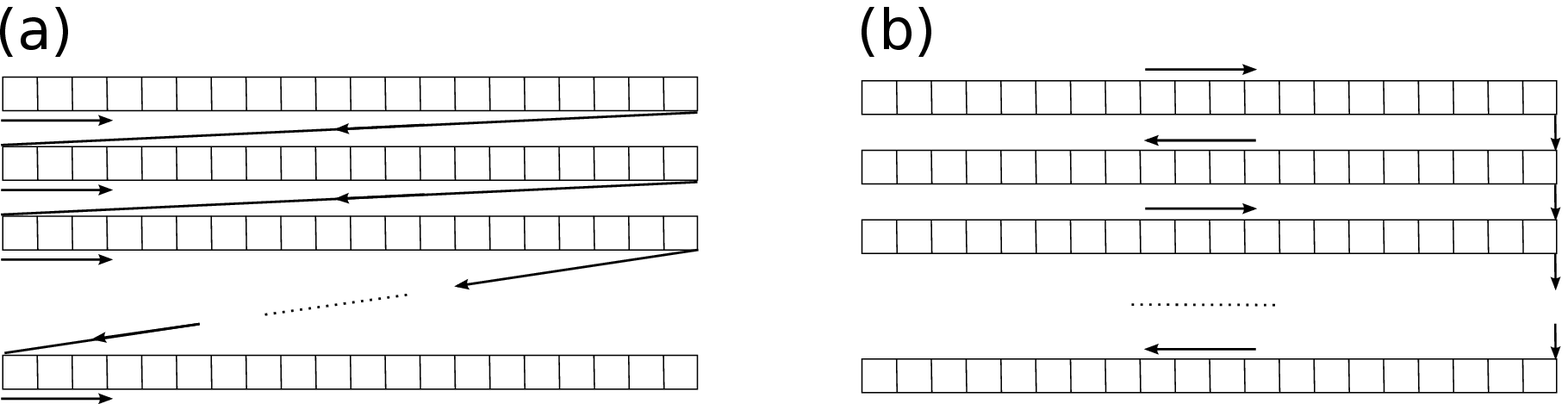}
\caption{Mapping of the data vector $\sigma_{1;n}$ to a 2D array. (a) Row scan from left to right
with breaks at fixed size. (b) Row scan with interleaved reverse scan directions.}\label{fig:rowscan}
\end{figure}

The simplest possible mapping of a string to a two dimensional (2D) array is to fix a length $ymax$ and scan the sequence left to right, making a row break at every multiple of $ymax$ as shown in figure \ref{fig:rowscan}a. The value $\sigma_{i}$ will be mapped to the array position given by $(i\; mod \; ymax, i+\lfloor i/ymax \rfloor)$ , where $\lfloor x \rfloor$ is the largest integer less or equal to x. If two neighboring values $\sigma_{i}$, $\sigma_{i+1}$ are not split by a row break ($i\; mod \; ymax \neq 0$), they will be nearest neighbors in the 2D mapping. Yet, from figure \ref{fig:rowscan}a, it is clearly seen that from the four nearest neighbors of any interior point in the 2D mapping only two of them will actually be consecutive values in the original $\varSigma$ sequence, while the other two (from the row above and below) can be far away in the $\varSigma$ sequence. A slight improvement to avoid the loss of locality at the row boundaries can be achieved by the scan shown in figure \ref{fig:rowscan}b, but the loss of locality from the above and below row remains a problem.

In any scan scheme, some loss of locality is unavoidable. In the 1D sequence each $\sigma_{i}$ will have two nearest neighbors, while in the 2D mapping, for any interior point, a coordination of four neighbors is achieved. So at least half of the nearest neighbors of $\sigma_{i}$ in the 2D mapping will not correspond to nearest neighbors in the 1D sequence. One will hope that all neighbors will still be close enough to the $\sigma_{i}$ in the $\varSigma$ string, and some meaningful clustering of close values is then achieved in the 2D mapping.

A useful magnitude for measuring locality is the square-to-linear ratio or dilation factor defined as  \cite{bauman06}:

\begin{equation}
 \Gamma(i,j)=\frac{d_{m}^2}{|i-j|}=\frac{|m(i)-m(j)|}{|i-j|}
\end{equation}

for a pair of points index $i$, $j$ and the mapping function $m:I\rightarrow I^2$ which maps the $\varSigma$ string to the unit square.

For each coordination distance $r=|i-j|$ in the $\varSigma$ string, the mean neighbor distance $\overline{d_{m}}$ in the 2D array is calculated from all interior points. The slower $d_{m}$ scales with increasing $|i-j|$ value, the better the locality preserving property of the mapping \cite{bauman06}. If the $d_{m}$ vs $|i-j|$ curve lies below the $d_{m}=|i-j|$ law we can consider the mapping to preserve locality fairly enough. 

\begin{figure}
\includegraphics[scale=0.7]{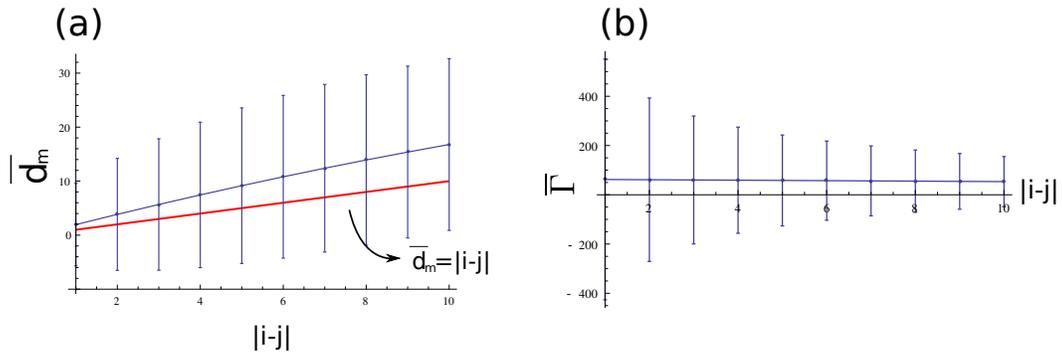}
\caption{Locality preserving property of the row scan. (a) Mean neighbor distance $\overline d_m$ as a function of
sequence distance $|i-j|$. The error bars correspond to the standard deviation showing a large dispersion of values
for all $\overline d_m$ values. (b) The mean dilation factor $\overline \Gamma$ has an almost constant behavior with
sequence distance $|i-j|$. A $\overline \Gamma$ value of around 60 is attained.}\label{fig:rowscan1}
\end{figure}

Figure \ref{fig:rowscan1}a shows that for the row scan, $d_{m}$ scales with $|i-j|$ worse than  $d_{m}=|i-j|$ curve. More importantly, the large dispersion of values around the mean value for each $|i-j|$, witness the fact that large variations in local neighborhood happens in this mapping. The same conclusion can be drawn from the dilation factor plot of Figure \ref{fig:rowscan1}b.  An almost constant behavior for $\Gamma(i,j)$ is observed around $60$.

The locality behavior we are searching for, is met (as best as possible) by the mapping known as Hilbert curve \cite{bauman06}, which will be explained in the next section.

\section{The Hilbert plot}\label{sec:hc}

Hilbert curve can be defined by a recursive algorithm. In two dimensions, the general idea is to divide, at step $k$, the unit square in four quadrants, and in each quadrant place a scaled down copy of the curve drawn in step $k-1$, in specified orientations. For each curve an entry and exit point is defined. Curves in adjacent quadrants are joined by exit to entry points. 

\begin{figure}
\includegraphics[scale=0.70]{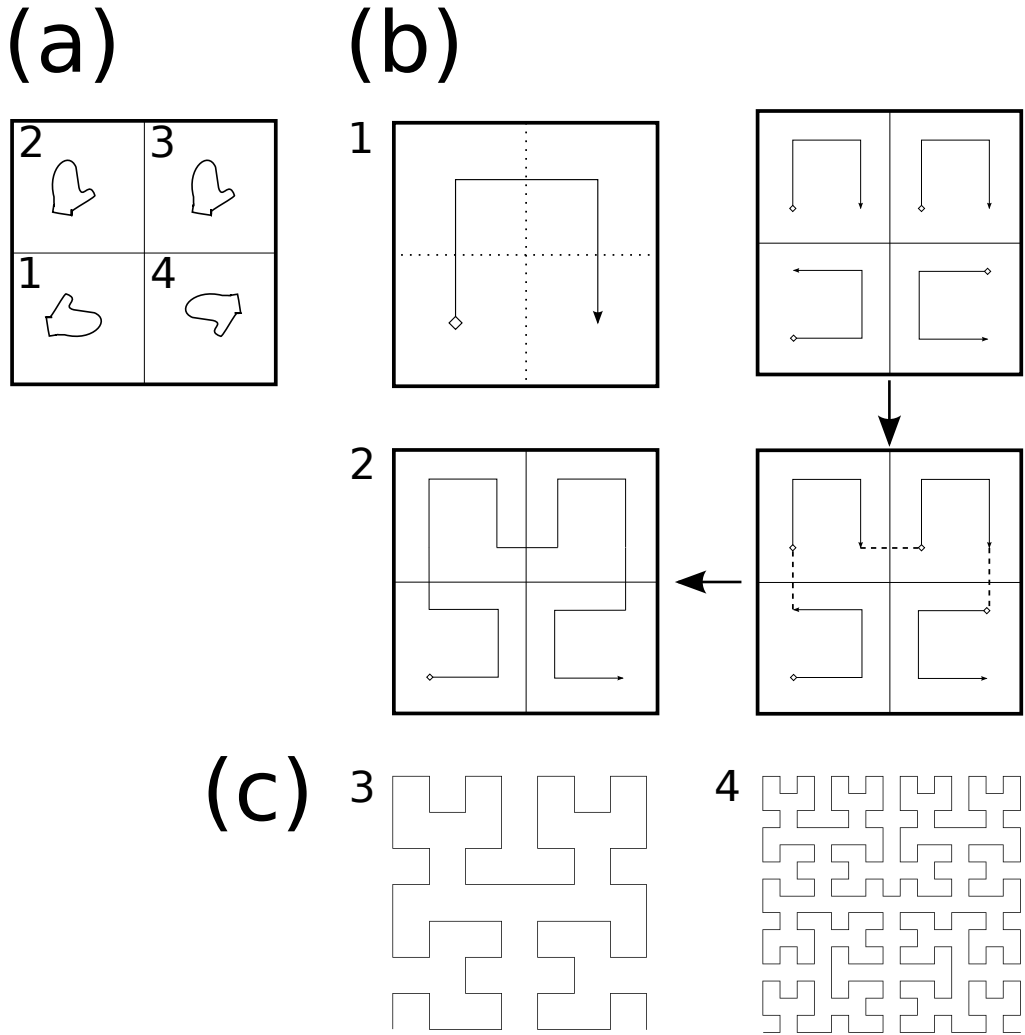}
\caption{Space filling Hilbert curve. (a) At any step $k$, the unit square is divided in quadrants and in each quadrant
scaled down copies of the curve drawn at step $k-1$ are drawn in different orientation. (b) Hilbert curve of
order $1$ is a staple like figure. At step $2$ the staple like curve of order $1$ is drawn at each
quadrant according to the rule depicted in (a). Exit points from quadrant $1$, $2$ and $3$ are linked with entry points
in quadrants $2$, $3$ and $4$, respectively, to make a continuous curve. (c) Hilbert curve of order
$3$ and $4$.}\label{fig:hchand}
\end{figure}

Let us numerate, starting from the lower left quadrant, all quadrants in clockwise direction (Figure \ref{fig:hchand}a). In the case of the Hilbert curve, in step $k$, the curve drawn in step $k-1$ is scaled down by $1/2$ and placed as it is in quadrant 2 and 3, while in quadrant 1 and 4, the curve is vertically flipped and rotated $+90^o$ (counterclockwise) and $-90^o$, respectively (Figure \ref{fig:hchand}a). 

Step by step, the algorithm is the following (we closely follow \cite{seebold07}):

\begin{enumerate}
 \item At step 1, the unit square contains the staple like curve depicted in Figure \ref{fig:hchand}b1.

 \item From step $k$ to $k+1$ the curve at step $k$ is scaled down by a factor of two and four copies are placed in each quadrant as already described above (Figure \ref{fig:hchand}b2).

 \item The curve is made continuous by joining the exit point of the curve in the first (respectively the second and third) quadrant with the entry point of the curve in the second (respectively the third and fourth) quadrant. The entry (exit) point of the new curve is the entry (exit) point of the copy placed in the first (fourth) quadrant (Figure \ref{fig:hchand}b2) 
\end{enumerate}

The recursive iteration of the above algorithm yields, at each step, a finer grained Hilbert curve $Hc(k)$ over the unit square (Figure \ref{fig:hchand}c). At step $k$, the Hilbert curve covers $4^k$ points in the square grid as a $2^k \times 2^k$ array.

The Hilbert mapping $\varSigma \otimes Hc(k)$, will associate to every index $i$ of the sequence $\varSigma$ a point $(x,y)$ over a square grid, such that the $(x,y)$  will have the value $\sigma_{i}$. The mapping $m:i \longrightarrow (x,y)$ will be done according to the path described by the Hilbert curve $Hc(k)$ of order $\lceil \log_{4}|\varSigma|\rceil$, where $\lceil x\rceil$ is the smaller integer larger than $x$. We will call such mapping a Hilbert plot of $\varSigma$. Several efficient algorithms have been reported to find the point in the Hilbert mapping from the index $i$, and viceversa (see for example \cite{breinholt98}).

\begin{figure}
\includegraphics[scale=0.7]{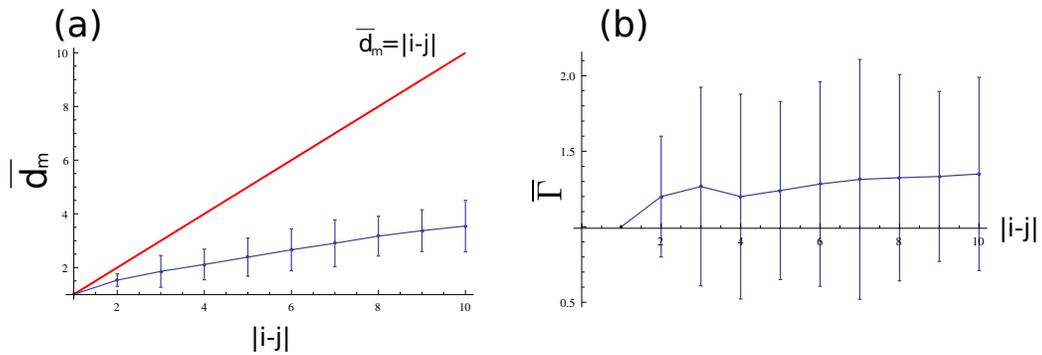}
\caption{Locality preserving property of the Hilbert curve (a) Mean neighbor distance $\overline d_m$ as a function of
sequence distance $|i-j|$. The scaling law is well below the $\overline d_{m}=|i-j|$ curve which can be considered
evidence of good neighbor preserving. The error bars correspond to the standard deviation. (b) The mean dilation factor
$\overline \Gamma$ behavior as a function of $|i-j|$.}\label{fig:hcdilation}
\end{figure}

Compared to the row scan, the Hilbert curve mapping has a $\overline{d_{m}}$ scaling with $|i-j|$ below the $d_{m}=|i-j|$ line, showing a much better locality preserving property (Figure \ref{fig:hcdilation}). Furthermore, the dispersion of values around the mean for each $|i-j|$, is significantly smaller (in relative terms). In correspondence with such behavior, the mean dilation factor is an order of magnitude below the one found for the row scan. Bauman \cite{bauman06} has demonstrated that for the Hilbert curve a upper bound for the dilation factor of 6 is achieved, which is considered almost optimal among all similar space filling curves.

From the construction of Figure \ref{fig:hchand}b, it can be seen that locality is mostly broken at the quadrants boundaries, specially between the first and fourth quadrant, special care must be taken when observing data across the quadrant lines.

In the next section the Hilbert plot will be used to visualize data from different sources.

\section{Hilbert curve visualization of 1D data sets}\label{sec:cases}

\subsection{Periodic sequences}

\begin{figure}
\includegraphics[scale=0.6]{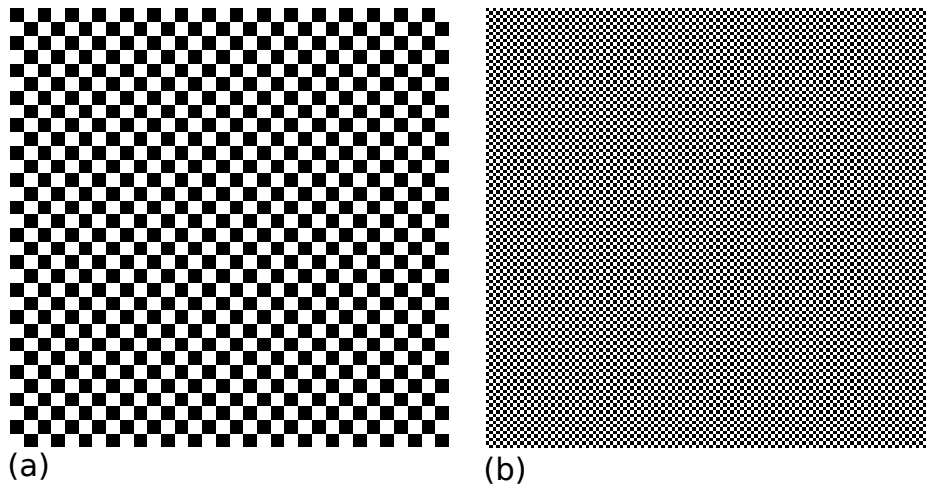}
\caption{ The Hilbert plot for the $\cdot 10\cdot$ periodic pattern, a sequence length of  (a) $4^5=(1024)$ and (b)
$4^7(=16 384)$ was taken.
}\label{fig:hcp}
\end{figure}

We start by considering the periodic pattern resulting from the repetition of the binary string $\cdot 10\cdot$. Figure \ref{fig:hcp} shows the Hilbert plot of such sequences for two lengths. A clear checkerboard pattern appears and it is preserved at any order. The Hilbert plot fits our intuitive idea of how such alternating sequence should look. Periodicity is clear from the plot. One must be cautious not to push the 2D interpretation to far. Connectivity in the original sequence is sequential, while the Hilbert plot suggest connectivity of similar values through the diagonals, this connectivity does not happen in the sequence.

\begin{figure}
\includegraphics[scale=0.8]{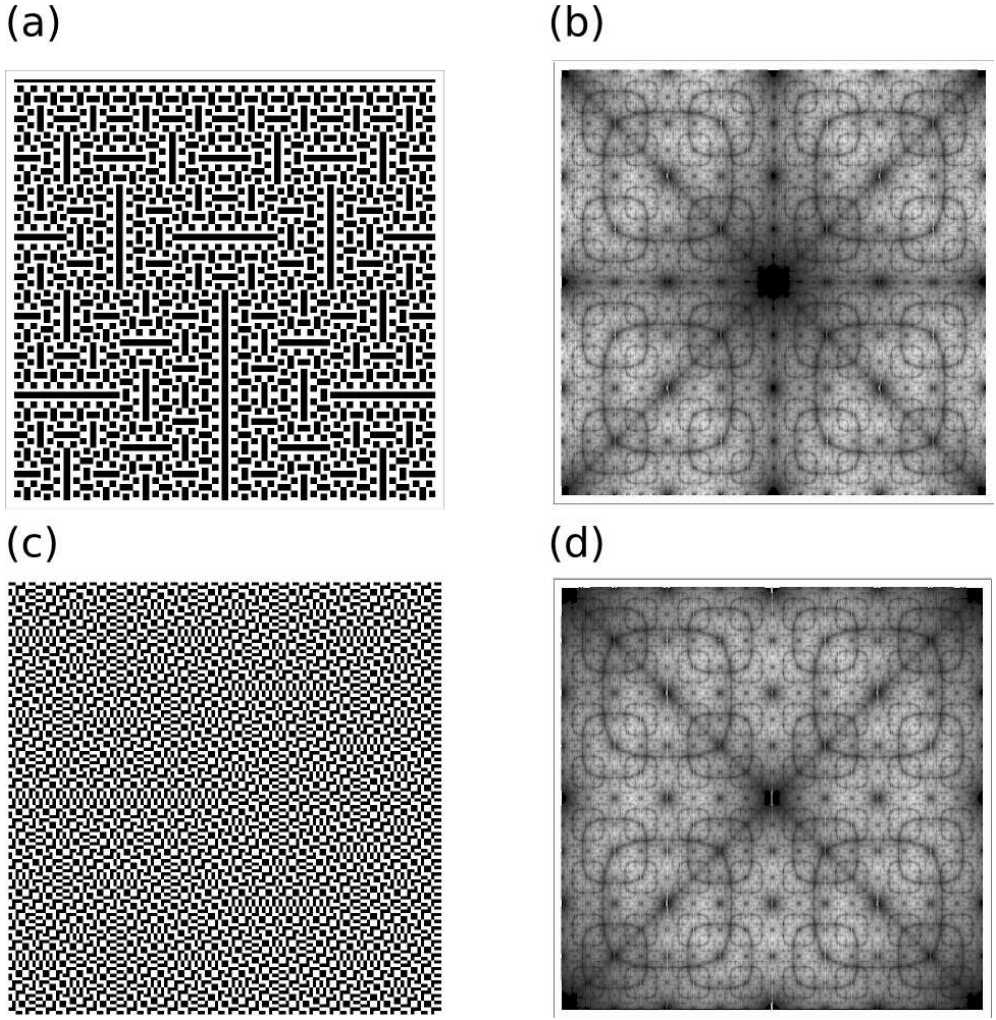}
\caption{The Hilbert plot and corresponding Fourier map for the (a) $\cdot 1001\cdot$  and (c)
$\cdot 1100\cdot$ periodic patterns. The Fourier map for both sequences are displayed in (c) and (d), respectively. For
both sequences a  length of $4^7(=16 384)$ was taken for the Hilbert Plot, the Fourier map was done
with a sequence of length $4^9(=262144)$.}\label{fig:1001}
\end{figure}

Not all periodic sequences give rise to easily identifiable patterns in the Hilbert plot. Figure \ref{fig:1001}a shows the Hilbert plot of the binary periodic pattern $\cdot 1001\cdot$. It is not immediately clear that such pattern corresponds to a periodic structure. The difficulty to identify a pattern is further emphasized if we consider the periodic sequence $\cdot 1100\cdot$ of the same length (Figure \ref{fig:1001}c). Both sequences are equivalent upon an odd cyclic shift and yet, their Hilbert plot looks different to a point, where the relation between both sequences can not be derived just by comparing both plots. If instead of looking at the Hilbert plot, we perform a Fourier transform of both arrays, the similarity between both sequences can be inferred from the similarity of their highly symmetric Fourier pattern (Figure \ref{fig:1001}b and d). It is clear, from both Fourier maps, that the sequence shows some sort of order. The Fourier patterns are each self similar and both exhibit a four fold symmetry axis and four mirrors, two cutting the image in quadrants and two along the diagonals. The overall symmetry of the patterns belongs to the $4mm$ point group \cite{mckie86}.

In any case the periodic nature of both sequences can be revealed if we look at the  4-granularity representation of both strings (not shown), a solid color Hilbert plot will emerge corresponding to the homogeneity of the 4-granularity representation. 

\begin{figure}
\includegraphics[scale=1]{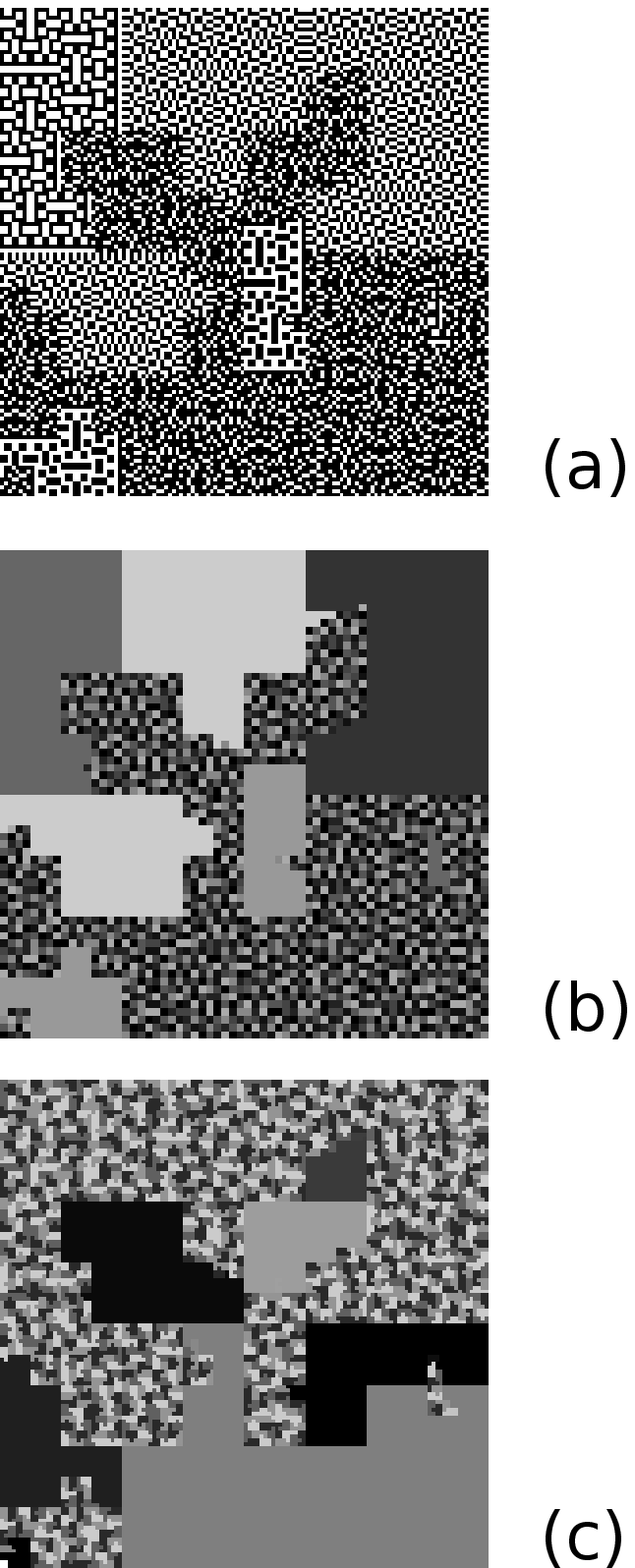}
\caption{Hilbert plot of a random interleaved sequence of $\cdot 1001\cdot $ and $\cdot 1101011\cdot $ periodic
intervals. (a) Faithful representation; (b) 4-granularity representation, solid blocks corresponds to the $\cdot
1001\cdot $ intervals; (c) 7-granularity, solid blocks corresponds to the $\cdot 1101011\cdot$ intervals. The length of
the sequence was taken as $4^{7}(=16384)$ for the Hilbert plot.}\label{fig:pergran}
\end{figure}

The usefulness of tuning granularity to exhibit periodic behavior in the data sequence can be seen more clearly when two or more periodic patterns are interleaved in the sequence. Consider a sequence made from the random interleaving of $\cdot 1001\cdot$ and $\cdot 1101011\cdot$ periodic ''chunks''. We consider the case where any periodic interval is not longer than $10^{-2}$ the total length of the sequence. Figure \ref{fig:pergran} shows the Hilbert plot of $\varSigma$ at three granularity levels: the faithful representation, the 4-granularity and the 7-granularity representation, the last two being the periodicity of the $\cdot 1001\cdot$ and $\cdot 1101011\cdot$ patterns respectively. 

In the Hilbert plot of the faithful representation  different regions can be identified but the periodic nature of each region is far from being clear. The Hilbert plot of the 4-granularity representation immediately identifies those regions with periodicity of the same value as solid blocks, while the corresponding Hilbert plot of the 7-granularity representation shows as one color blocks the complementary regions corresponding to the $\cdot 1101011\cdot$ periodic ''chunks''. The three plots together allow to infer that the $\varSigma$ sequence is formed by intervals of periodicity 4 and 7, and no other intervals are present. The fact that the periodic intervals of the $\varSigma$ string show up in the Hilbert plots as clearly identifiable regions, points to the convenience and importance of the clustering and locality preserving properties of the Hilbert curve.

\subsection{Aperiodic sequences}\label{sec:norm}

Next, we will consider aperiodic fully ordered sequences. The Thue-Morse sequence and the Fibonacci sequence are non periodic, rule based with entropy rate equal\footnote{Entropy rate, is a length invariant measure of the amount of new information gained per unit time in a dynamical process \cite{cover06}} zero \cite{lothaire02}. The Thue-Morse sequence is defined by the morphism

\begin{eqnarray*}
tm(0)=01\\
tm(1)=10
\end{eqnarray*}

while the Fibonacci sequence follows the rule

\begin{eqnarray*}
fb(0)=01\\
fb(1)=0
\end{eqnarray*}

Starting with a value of $1$, the first characters  will be

$1001011001101001011010011001011001101001100101101$

for the Thue-Morse morphism, and 

$1011010110110101101011011010110110101101011011010$

for the Fibonacci sequence.

\begin{figure}
\includegraphics[scale=0.7]{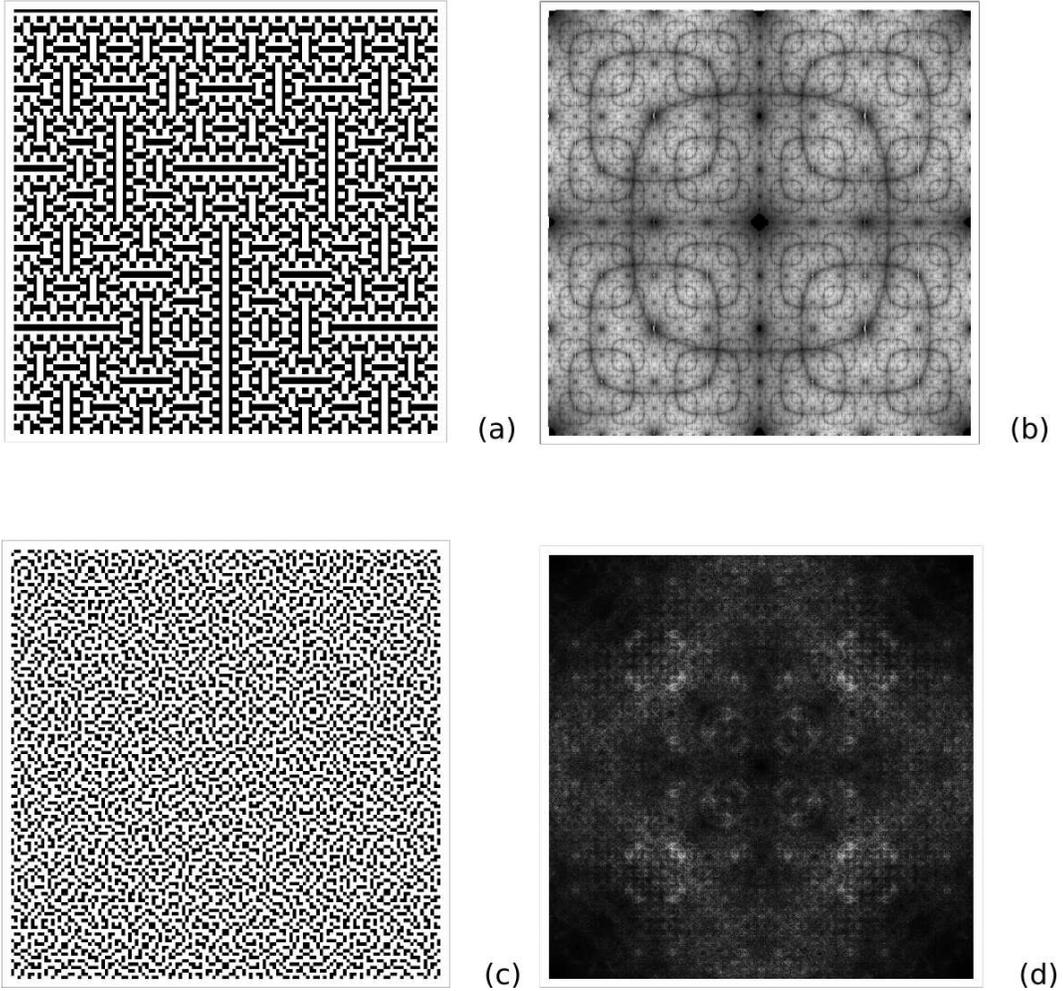}
\caption{Hilbert plot of the (a) Thue-Morse and (c) Fibonacci sequences. The
Fourier pattern of the Thue Morse sequence (b) is highly symmetric and self similar, while for the
Fibonacci sequence (d), a two fold symmetry is observed. For
both sequences a  length of $4^7(=16 384)$ was taken for the Hilbert Plot, the Fourier map was done
with a sequence of length $4^9(=262144)$.}\label{fig:tmf}
\end{figure}

Figure \ref{fig:tmf} shows the Hilbert plot for both sequences, together with its Fourier pattern. Both Hilbert plots show some type of structure, being more visually prominent in the case of the Thue-Morse sequence. The Fourier pattern of the Thue-Morse Hilbert plot exhibits self similarity which can be seen at the $2^{-1}$ (quadrants), $2^{-2}$ (octant) and $2^{-3}$ scales. The symmetry of the pattern is again $\overline{4}mm$.

The Fourier pattern of the  Fibonacci Hilbert plot is also symmetric. In this case, the pattern has a two fold axis together with two mirror planes perpendicular to each other and dividing the square into quadrants. The point group symmetry of the pattern is also $4mm$.

\subsection{Logistic Map}

We now turn to the logistic map \cite{medio03} given by the equation 

\begin{equation}\label{eq:lmap}
 x_{n+1}=1-r x_{n}^2 
\end{equation}

A binary generating partition at $x=0$ is used to reduce the output to a binary sequence. Different behaviors are obtained by changing the $r$ parameter: at $r=1.8$, chaotic behavior is observed, with an entropy rate of $h=0.5828$; at $r=1.7499$, a strong intermittent point is observed and an entropy rate of $h=0.2597$ is achieved; and at $r=1.40115518$ the map is at the Feigenbaum point, where the entropy rate becomes zero.

\begin{figure}
\includegraphics[scale=0.75]{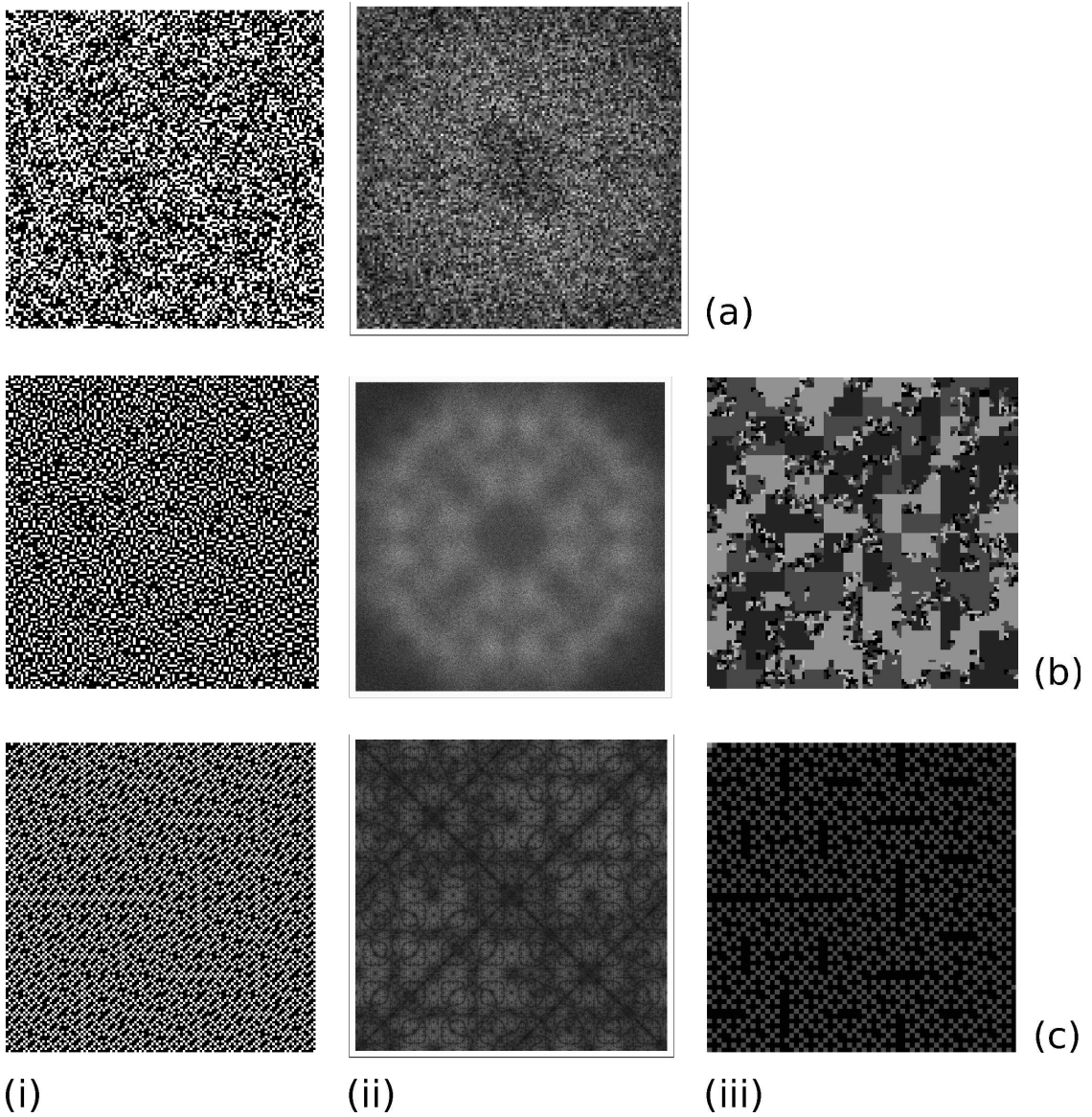}
\caption{Hilbert plot (i) and the corresponding (ii) Fourier map for the logistic map at (a) r=1.8 (in the chaotic
region); (b) r=1.7499 (an intermittent point) and; (c) the Feigenbaum point (with vanishing entropy rate). The (iii) column shows
the Hilbert plot for (b) 3-granularity representation of the intermittent point and, (c) the 4-granularity
representation of the Feigenbaum point.}\label{fig:logmap}
\end{figure}

Figure \ref{fig:logmap} shows the Hilbert plot for each $r$ value. Chaotic behavior at $r=1.8$, gives rise to a random looking Hilbert plot (Figure \ref{fig:logmap}a) with no special feature, the Fourier pattern also fails to show any structure. Visualization of the Hilbert plot for increasing granularity, did not improve the featureless character of the Hilbert plot. Is hard to appreciate if the Hilbert plot at the intermittent point shows any pattern, for the faithful representation of the sequence (Figure \ref{fig:logmap}b), a fact that is not clearly elucidated by the Fourier pattern. Yet the Hilbert plot of the 3-granularity representation of the sequence immediately shows, that in fact, the sequence has mostly a structure compatible with a pattern with a periodicity of 3, interleaved with some minor random subsequences, which accounts for the entropy rate of $0.2597$. The inspection of the sequence reveals  $\cdot 110\cdot$ as the periodic pattern. At the Feigenbaum point ($r=1.40115518$), the Hilbert plot (Figure \ref{fig:logmap}c) clearly shows the emergence of structure, which is also revealed by the Fourier pattern. The Hilbert plot of the 4-granularity representation almost shows a checkerboard pattern, a look into the sequence allows to identify two recurring patterns: $\cdot 1110\cdot$ and $\cdot 1010\cdot$, yet their repetition is not periodic. 

\subsection{Ising nearest neighbor model}\label{sec:comparison}

We also considered the spin-$1/2$, nearest-neighbor Ising model, the reader is referred to reference  \cite{feldman98} for detailed discussion of the model. For our purposes, it will be enough to consider three regions: (a) Close neighbor interaction prevails over the thermal energy, and leads to the ferromagnetic regime; (b) close neighbor interaction prevails over the thermal energy, and leads to the antiferromagnetic state; (c) thermal
energy is the most important contribution and the state is paramagnetic.

\begin{figure}
\includegraphics[scale=0.40]{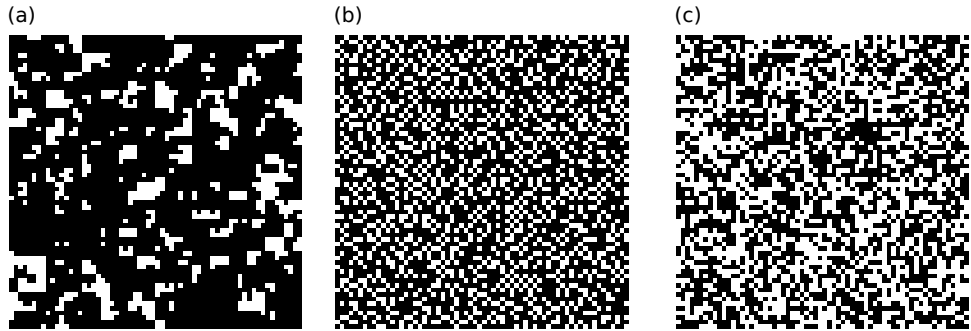}
\caption{Hilbert plot of the 1/2-Ising nearest neighbor model in the (a) ferromagnetic, (b)
antiferromagnetic and (c), paramagnetic state.}\label{fig:ising}
\end{figure}

The Hilbert plot of the three states is shown in figure \ref{fig:ising}. The black color corresponds to a spin orientation (up spin) while the white color  is then associated with the opposite direction (down spin). Ferromagnetic state is characterized by large regions with the same spin orientation (black regions) corresponding to long intervals of the same spin orientation, typical of such state. Isolate spin orientation are rare. Care again must be taken, not to take the interpretation of the Hilbert plot to far away. Although it may seem that the black regions are connected, the
actual 1D sequence does not exhibit such connectivity as spin runs along a single direction.

The antiferromagnetic state exhibits a Hilbert plot where roughly the same amount of up and down spins are observed in a pepper and salt type of image. Isolated spins in both directions can be commonly seen. 

If compared to the previous states, the Hilbert plot of the disordered state has a more random looking plot, with small regions of both up and down spin failing to show any structure. The Fourier pattern of all three states fail to show any significant structure resulting from the lack of long range order in all cases.

We further used the Hilbert plot visualization to identify  phases in Monte Carlo simulations of polytypes in layered structures using Ising type models (not shown). The visualization techniques allowed in al cases the preliminary identification of the emerging polytypes even when interdispersed with shrinking ones. Results will be published elsewhere.

\subsection{Real data example}\label{sec:realdata}

\begin{figure}
\includegraphics[scale=0.50]{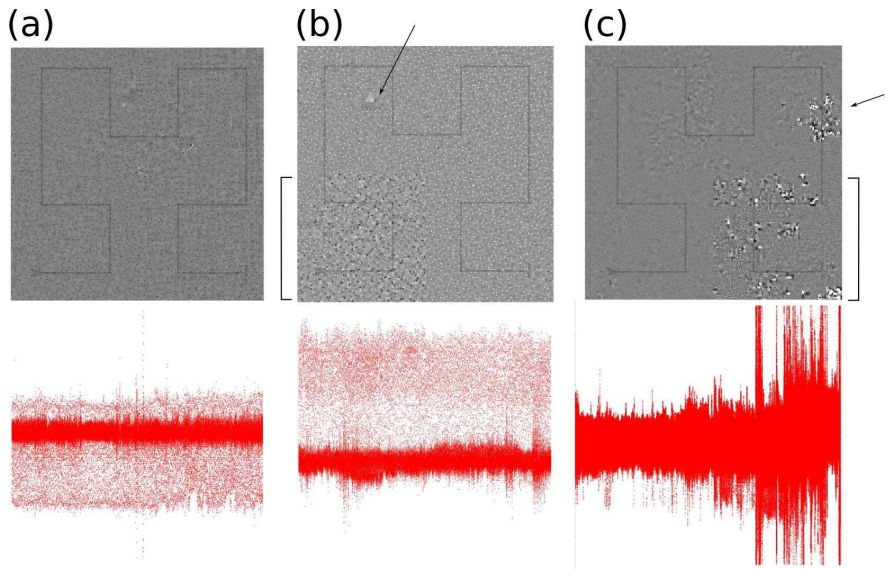}
\caption{Hilbert plot (above) and scatter plot (below) of real electrocardiogram of a (a) healthy person, (b) patient with supraventricular arrhythmia and (c), patient with severe heart condition leading to sudden death. All data corresponds to a 24 hour time span, 1 second resolution. the number of points in each plot is above $3\;10^6$. Over imposed Hilbert curve aids in following the flow of data. See text for details.}\label{fig:heartbeat}
\end{figure}

We have been studying the long time correlation of heart activity based on a previous model forwarded by Peng et al\cite{peng93}. In their work, heartbeat is treated as a (possible) correlated signal and it is found, that in large time scales (24 hours), the existence of certain long correlations could be linked to the heart condition. As a result of the study, it was also found that the heartbeat could be possible modeled as a random walk. We have extracted data from the PhysionBank archives which contains physiologic signals and related data for use by the biomedical research community, including heart activity\cite{physio}. In figure \ref{fig:heartbeat} (Higher resolution images can be found as supplementary files), the Hilbert plot, together with the scatter plot of the data for a healthy person, a patient with arrhythmia and a patient that suddenly died of heart attack while monitoring, are shown. The data sets are each above 3 million points and spans around 24 hours of heart activity. A Hilbert curve is over imposed the plot to aid in following the time flow. 

Hilbert plot is able to show difference in all cases, while giving a level of detail not clearly appreciated in the scatter plot. The healthy person shows a homogeneous, mostly uniform plot, while in the person with arrhythmia, a more ''salt and pepper'' type of image is seen. The plot is not completely homogeneous, as the ``salt and pepper'' seems to be heavier at the first half of the plot. A single heart event is seen marked by arrow, which was identified at around 10 hours of monitoring.  For the patient that underwent a heart attack, a first episode happens at around 12 hours of monitoring (pointed by arrow). In the Hilbert plot we can see details of this first event as different gray levels, which is lost in the scattered plot due to the cluttering of the points. From this first event the patient recovered, and shortly after a successive string of heart events, lead to the final stroke. Again, details within this successive events can be qualitatively be seen in the Hilbert plot, while is difficult in the scattered plot.  Hilbert Plot of the segmented data in chunks of $10^6$ points, allow to observe even finer details in the heart activity for 8 hours time span (not shown).

Hilbert plot allows to see in the same plot scale, the overall picture of the 24 hours, with all heart events and their possible correlation, while not loosing details due to cluttering of the data points. Hilbert plot Fourier transform showed peaks revealing the underlying pattern of the heart beats (not shown).

\section{Conclusions}

A definite solution to the visualization of large data series which is completely satisfactory, remains elusive and is strictly unreachable. This points to the need of making a good compromise between a visualization that allows the identification of features and regularities and at the same time, avoids artificial artifacts. Hilbert plots proves to strike a good balance between both extremes.  

It is important to bare in mind that in observing Hilbert plots and its Fourier map, one must be careful not to stretch the interpretation of the visualization to far, and infer conclusions that can be the result of different topologies from the 1D and 2D spaces, a problem common to any other visualization. On the other hand, the clustering property of the Hilbert plot, and its good neighborhood preserving property, makes it useful in the visualization task. A price must be paid though, it is not easy to imagine the sequence path in a dense enough Hilbert plot. Yet, one gets, for such price, the clustering in connected regions of neighboring values in the data vector. Additionally, the use of granularity representation allows to observe the data at different levels of detail and to identify periodic or nearly periodic intervals in a very straightforward manner.

It is important to realize that a single visualization technique proves almost impossible to capture all the visual information of the data. The Fourier map of the Hilbert plots seems to grasp complementary visual information not easily seen in the Hilbert plot. In particular, it captures the  presence of structure and order, being periodic or aperiodic.

Finally, the availability of modern computer tools allows to use any visualization in a dynamical manner, making it possible to fully exploit the strength of a particular representation. In the case of the Hilbert plots, for example, the difficulties in associating a particular point of the 2D representation with its index in the data sequence, can be easily solved. Other facilities such as zooming, image segmentation and simultaneous display of different granularity representation, among others, can also be easily implemented by software.

\section{Availability}

Software used for the Hilbert plots shown is freely available upon request from the authors.

\section{Acknowledgments}

This work was partially financed by FAPEMIG under the project BPV-00047-13 and BPV-00039-12. EER which to thank PVE/CAPES for financial support under the grant 1149-14-8. 


\pagebreak

\end{document}